\documentclass[useAMS,usenatbib,10pt]{mn2e} 
\usepackage{graphicx}
\usepackage{txfonts}

\def\amin{$^{\prime}$\/\ }

\def\HI{H\,{\sc i}}
\def\smh{$M_{\rm H_{I}}/D_{l}^{2}$}
\def\mh1{$M_{\rm H_{I}}$}

\def\kms{km s$^{-1}$}
\def\hdef{$def_{\rm H_{I}}$}
\def\etal{{\it et al.}\thinspace}

\def\ctab#1#2#3#4{
\begin{table}
 \begin{minipage}{#1 mm}
 \label{#2}
 \caption{#3}
 {#4}
\end{minipage}
\end{table}
}

\def\cfig#1#2#3#4#5#6#7{
    \begin{figure}
    \hspace{#7cm}
    \centerline{\rotatebox{#6}{\includegraphics*[height=#3in]{#1}} }
    \vspace*{#5in}
    \caption{#2}
    \label{#4}
    \end{figure}
}

\title[\HI ~ content in galaxies in loose groups]{\HI ~ content in galaxies in loose groups}
\author [Sengupta, C. \& Balasubramanyam, R.]{Chandreyee Sengupta$^{1}$\thanks{e-mail:csg@rri.res.in} \& Ramesh Balasubramanyam$^{1}$\thanks{e-mail:ramesh@rri.res.in}\thanks{Send offprint requests to: Chandreyee Sengupta} \\
$^{1}$Raman Research Institute, Bangalore 560 080 INDIA}

\begin{document}
\date{}
\pagerange{\pageref{firstpage}--\pageref{lastpage}} \pubyear{0000}

\maketitle

\label{firstpage}

\begin{abstract}
\noindent Gas deficiency in cluster spirals is well known and ram-pressure stripping is considered the main gas removal mechanism. In some compact groups too gas deficiency is reported. However, gas deficiency in loose groups is not yet well established. Lower dispersion of the member velocities and the lower density of the intra-group medium in small loose groups favour tidal stripping 
as the main gas removal process in them. Recent releases of data from \HI ~Parkes all sky survey (HIPASS) and catalogues of nearby loose groups with associated diffuse X-ray emission have allowed us to test this notion. In this paper, we address the following questions: (a) do galaxies in groups with diffuse X-ray emission statistically have lower gas content compared to the ones in groups without diffuse X-ray emission? (b) does \HI ~deficiency vary with the X-ray luminosity, $L_{x}$, of the loose group in a systematic way? We find that (a) galaxies in groups with diffuse X-ray emission, on average, are \HI ~deficient, and have lost more gas compared to those in groups without X-ray emission; the later are found not to have significant \HI ~deficiency; (b) no systematic dependence of the \HI ~deficiency with $L_{x}$ is found. Ram pressure assisted tidal stripping and evaporation by thermal conduction are the two possible mechanisms to account for this excess gas loss.
\end{abstract}


\section {Introduction}
In field galaxies, neutral hydrogen gas gets either converted into molecular form (and into stars) or ionised. A small fraction does escape in galactic winds. However in galaxies in clusters, substantial amount of gas goes into the intra-cluster medium (ICM), making the members {\it deficient} in \HI ~relative to the gas content of a field galaxy of a similar morphological type.  Such \HI ~deficiency in cluster spirals is well reported in the literature. Members in groups may also lose gas to the intra-group medium (IGM) and become relatively {\it deficient} in \HI. Galaxies in some of the Hickson compact groups have been reported to be gas deficient \citep{H1HC1} but such deficiency is still debated \citep{H1HC2}.
 Gas deficiency in loose group environment has not yet been investigated systematically. For the first time, \HI ~deficiency by a factor more than 1.6 has been reported in some members of a non-compact group in the Puppis region
\citep{H1NC1}. 
In some of the loose groups, the IGM is found to have enhanced metalicity, suggesting recent gas removal from the member galaxies \citep{TRamP}.
Thus, investigating gas removal processes operating in loose groups is timely and important.
\vspace*{0.3cm}

Two mechanisms are considered important for gas removal: tidal interaction and ram-pressure stripping.
Ram pressure stripping \citep{RamP}
is effective when the \HI ~surface mass density is less than
$\rho_{0}{\rm v}^{2}/(2\pi~G\sigma_{*})$, where $\sigma_{*}$ is the stellar surface mass density.
Clearly, larger the ICM or IGM density, $\rho_{0}$, and galaxy
velocity dispersion, ${\rm v}$, the more effective is this stripping. Clusters satisfy these requirements
and ram-pressure stripping is considered an effective process in them. In groups, though, both these
quantities are lower, especially dispersion by a factor of 10 and therefore ram pressure stripping is
considered ineffective. Tidal interactions on the other hand, involve gravitational interaction
between two or more galaxies as they pass by each other. The lower relative velocities in groups allow
larger interaction timescales making tidal stripping the likely process for gas removal.
This, in a larger sense, includes galaxy `harassment',
where the tidal stripping is caused by the overall gravitational potential of the group. However, a significant number of loose groups have been found to have hot gas in them emitting diffuse X-rays \citep{DifX}.
In such cases, ram pressure assistance cannot be ruled out, making X-ray bright groups more \HI ~deficient than non X-ray bright groups.

\vspace*{0.3cm}
\cite{H1NC1} studied selected galaxies in five groups, NGC 533, 5044, 2300, 5846 and 4261, detected to have a hot IGM by ROSAT, and found a few of them to be strongly deficient in \HI. They find the deficiency to be related to the physical proximity of the galaxy to the X-ray region. We have undertaken to study in detail the \HI~ content of galaxies in groups with and without X-ray detection, covering a wide-range in X-ray luminosity.
Here we report the first part of our study, viz. a comparative analysis of \HI~ content in small loose groups based on HIPASS and other existing single dish  data. The paper is organised as follows: the next section deals with the sample selection, \HI~ data and its processing details. In Section 3, we report the analyses of the data and our results. The possible explanations of the results are discussed in Section 4. The work is summarised in Section 5.

\section {Sample, Data \& Processing}
 
Our sample is made of twenty seven groups, ten belonging to the X-ray bright category and seventeen belonging to the non X-ray bright category. 
We chose all the X-ray bright groups, totalling 10, from the X-ray atlas of nearby poor groups \citep{DifX} satisfying the following criteria: most members have single dish \HI~ measurements, their distances $\la$ 50 Mpc (a value of 100 \kms ~Mpc$^{-1}$ is used for the Hubble constant throughout this paper) and their membership is less than 25 (to avoid poor clusters). Their X-ray luminosities are in the range from  2.0$ ~\times 10^{40}$ and 6.3$ ~\times 10^{42}$ erg$~s^{-1}$. Eight X-ray non-detected groups (six from \cite{DifX}, and two from \cite{DifX1}) and 9 all-spiral groups (all the group members being late type spirals), make up the non X-ray bright category. Non compact all spiral groups are not expected to have diffuse X-rays from their IGM (\citealt{DifX}, \citealt{henry}, \citealt{ota}, \citealt{DifX1}) and therefore have been included to augment the non X-ray bright sample. The distances and memberships of these seventeen groups also follow the same criteria as the X-ray bright groups.

Among the 27 groups that make up our sample, 13 are from the southern hemisphere and 14 are from the northern hemisphere. For most of the southern hemisphere galaxies HIPASS database has been used. Given that typical \HI ~masses in these galaxies will be a few times $10^{8}$~$M_{\odot}$, the HIPASS sensitivity constrains that these groups be nearer than $\sim$ 40 Mpc. For the rest of the galaxies, \HI ~measurements from the literature have been used (\cite{H1-R1}, \cite{H1-R2}, \cite{H1-R3}, \cite{H1-R4}, \cite{H1-R5}, \cite{H1-R6}, \cite{H1-R7}, \cite{H1-R8}, \cite{H1-R9}). This imposes some practical difficulties: different surveys have different detection limits and resolution; some of the group members did not have any \HI~ observation; in some other cases, the errors on the integrated flux densities were not quoted and we have used an average of the errors quoted for the other group members. 
 
For group memberships we have followed the Lyon Group of Galaxies (LGG) catalog \citep{LGGC}. All LGG members of a group within a diameter of 1.2 Mpc (typical crossing distance in $\sim$ 5 Gyr) about the group center are included. In most cases, this circle seems a natural boundary. In a few cases where a significant fraction of LGG members lie outside this range, we have extended the circle to a diameter of 1.5 Mpc about the group center. This is done to ensure that the computed \HI~deficiency truely reflects the property of the group. Galaxies lying outside a diameter of 1.5 Mpc have been excluded.
We also include as members of the group all galaxies found using NASA Extragalactic Database (NED) within the spatial and velocity extent defined by the LGG members. 
All members thus chosen are tabulated in Tables 3 and 4 under their respective group names with distances to the groups derived from their velocities given in brackets. The table lists the source names, morphological type ($\it MT$), optical diameter in arcminute ($d_{l}$(\amin)), gas surface matter density in logarithmic units ($\it SMD$), \HI~deficiency (\hdef), angular distance from the group centre in arcmin (Ang.) and the telescopes used for obtaining the \HI~data (Tel.). Footnote at the end of Table 4 explains the symbols used for the different telescopes. Groups have been ordered in increasing right ascension but the galaxies within the groups have been ordered in increasing distance from their centers. The later ordering has been done to highlight the location of early-type and \HI ~deficient galaxies with respect to the group centers. For galaxies that 
belong to an early morphological type (E,S0 and S0/a), a ``$-$'' is marked in columns 4, 5 and 7; such galaxies have not been used in this study.  Spiral galaxies which do not have \HI ~data are denoted by an ``X'' in column 7. A ``ND'' in column 7 indicates a spiral not detected in \HI. Optical diameter, d$_{l}$, is taken from RC3 catalogue using NED: it is the optical major isophotal diameter measured at or reduced to a surface brightness level m$_B$ = 25.0 B-m/ss. 

The HIPASS spectra obtained towards the galaxies were used to find the centroid velocities and the integrated flux densities ($\int SdV$ in Jy\,km~s$^{-1},~S$ being the flux per beam per channel and $dV$ being the velocity resolution) after fitting and removing second order polynomial baselines. 
Unipops package was used for this processing. 
There were several instances of confusion in the case of HIPASS data (Parkes beam has an FWHM of 14' at 21 cm).
In the case of northern groups, multiple measurements with different telescopes  
were available. This and the fact that the beamsizes were smaller than that of Parkes reduced the instances of confusion. In all cases of confusion, the deficiencies have been calculated assuming it to be equal among the confused galaxies.

The total gas mass (\mh1) can not be straight away used for studying gas removal in galaxies as the \HI ~content depends both on their sizes and morphological types. In fact, the disk size seems to be a more important diagnostic for the \HI~ mass than the morphological type \citep{H1Def}. Gas mass surface density \smh ~proves to be a better measure of \HI ~content as it incorporates the diameter of the spiral disk. Another advantage of using \smh ~over \mh1 as a measure of \HI~ content is that \smh ~is distance independent and therefore free from its uncertainty.

\setcounter{table}{0}
\ctab{85}{tab1}{Expected surface matter densities for different morphological 
types (adapted from \citealt{H1Def} for using RC3 diameters instead of UGC diameters and taking h=1).
}{
\begin{center}
\begin{tabular}{p{4.3cm}r}
\hline
Morphological type ({\it M.T.})~ \& \hfill Index &log($\frac{M_{H_{I}}/D_{l}^{2}}{M_{\odot}/kpc^{2}}$)$_{pred} ~\pm$ s.d  \\ \hline 
Sa,Sab \hfill 2 &6.77 $\pm$ 0.32 \\ 
Sb \hfill 3 &6.91 $\pm$ 0.26 \\ 
Sbc \hfill 4 &6.93 $\pm$ 0.19 \\ 
Sc \hfill 5 &6.87 $\pm$ 0.19 \\ 
Scd,Sd,Irr,Sm,Sdm, dSp \hfill 6 &6.95 $\pm$ 0.17 \\ 
Pec \hfill 7 &7.14 $\pm$ 0.28 \\ \hline 
\end{tabular}
\end{center}
}

\par The expected values of 
 \smh ~for various morphological types are taken from \cite{H1Def}. While \cite{H1Def} used the UGC blue major diameters, we have used the RC3 major diameters. To take care of the difference in the surface matter densities that result from the use of RC3 diameters, we add a value of 0.08 \citep{RC3UGC} to the expected surface matter densities given by \cite{H1Def}. The final values of expected \smh~ used in this paper are given in Table 1.
\HI~deficiency of a single galaxy is determined following the usual definition viz.:
\begin{equation}
{\it def_{\rm {H_{I}}}=log{{\frac{M_{H_{I}}}{D_{l}^{2}}}|_{pred}}~-~log{{\frac{M_{H_{I}}}{D_{l}^{2}}}|_{obs}}} 
\end{equation}

An average of this over all the \HI~detected galaxies is used as a measure of the \HI~ deficiency of the groups, hereafter referred to as the ``group deficiency''.
Table 2 lists the group names, their X-ray luminosity in logarithmic units, the calculated group deficiencies with 1  $\sigma$ error on them, for the 10 X-ray bright groups, the 8 X-ray non-detected ones and the 9 all-spiral groups that have no reported X-ray observation.

\ctab{115}{tab2}{Estimated group deficiencies}{
\begin{tabular}{llr}
\hline
Group Name &$L_{x}$, erg s$^{-1}$ & Group deficiency \\ 
\hline
NGC524  &40.53 & 0.62 $\pm$ 0.240 \\
NGC720 &40.86 & 0.42 $\pm$ 0.138\\ 
NGC1589 (NGC1587) &40.92 & 0.46  $\pm$ 0.157\\
NGC3686 (NGC3607) &40.53 & 0.07 $\pm$ 0.062 \\
NGC4261 &41.89 & 0.16 $\pm$ 0.134  \\
NGC4589 (NGC4291) &40.74 & 0.12 $\pm$ 0.062 \\
NGC5044 &42.81 & 0.34 $\pm$ 0.082 \\ 
UGC12064 &42.17 & 0.06 $\pm$ 0.055 \\ 
IC1459 &40.52 & 0.29 $\pm$ 0.093 \\ 
NGC7619 &42.05 & 0.20 $\pm$ 0.053 \\
NGC584 &$<$40.51 & 0.36 $\pm$ 0.063\\
NGC1792 (NGC1808) &$<$39.79 & 0.20  $\pm$ 0.059\\
NGC5061 (NGC5101) & $<$40.77 &0.08 $\pm$ 0.102\\
NGC5907 (NGC5866) &$<$39.48 &-0.08 $\pm$0.174\\
UGC9858 (NGC5929) &$<$40.50 &0.39 $\pm$0.055\\
NGC7448 &$<$40.40 &-0.11 $\pm$0.050\\
NGC7582 &$<$40.74 &0.27 $\pm$ 0.055\\
NGC7716 (NGC7714) &$<$39.72 & 0.22 $\pm$ 0.032\\
NGC628 &no observation &-0.20 $\pm$0.154\\
NGC841&no observation&0.14 $\pm$0.088\\
IC1954 &no observation & 0.20  $\pm$ 0.108\\
NGC1519 &no observation & 0.24 $\pm$ 0.093\\
NGC2997 &no observation &0.10 $\pm$ 0.050\\
NGC3264 &no observation &-0.16  $\pm$0.060 \\
NGC4487 &no observation & 0.19  $\pm$ 0.104\\ 
NGC6949&no observation &-0.16 $\pm$0.139\\
UGC12843 &no observation &-0.05 $\pm$0.115 \\
\hline
\end{tabular} \\
Note: The group names given are from LGG catalogue. Those given in\\
brackets are from \citealt{DifX} and \citealt{DifX1}.
}

\section {\HI ~content in galaxies in groups with and without diffuse X-ray emission}	
\cfig{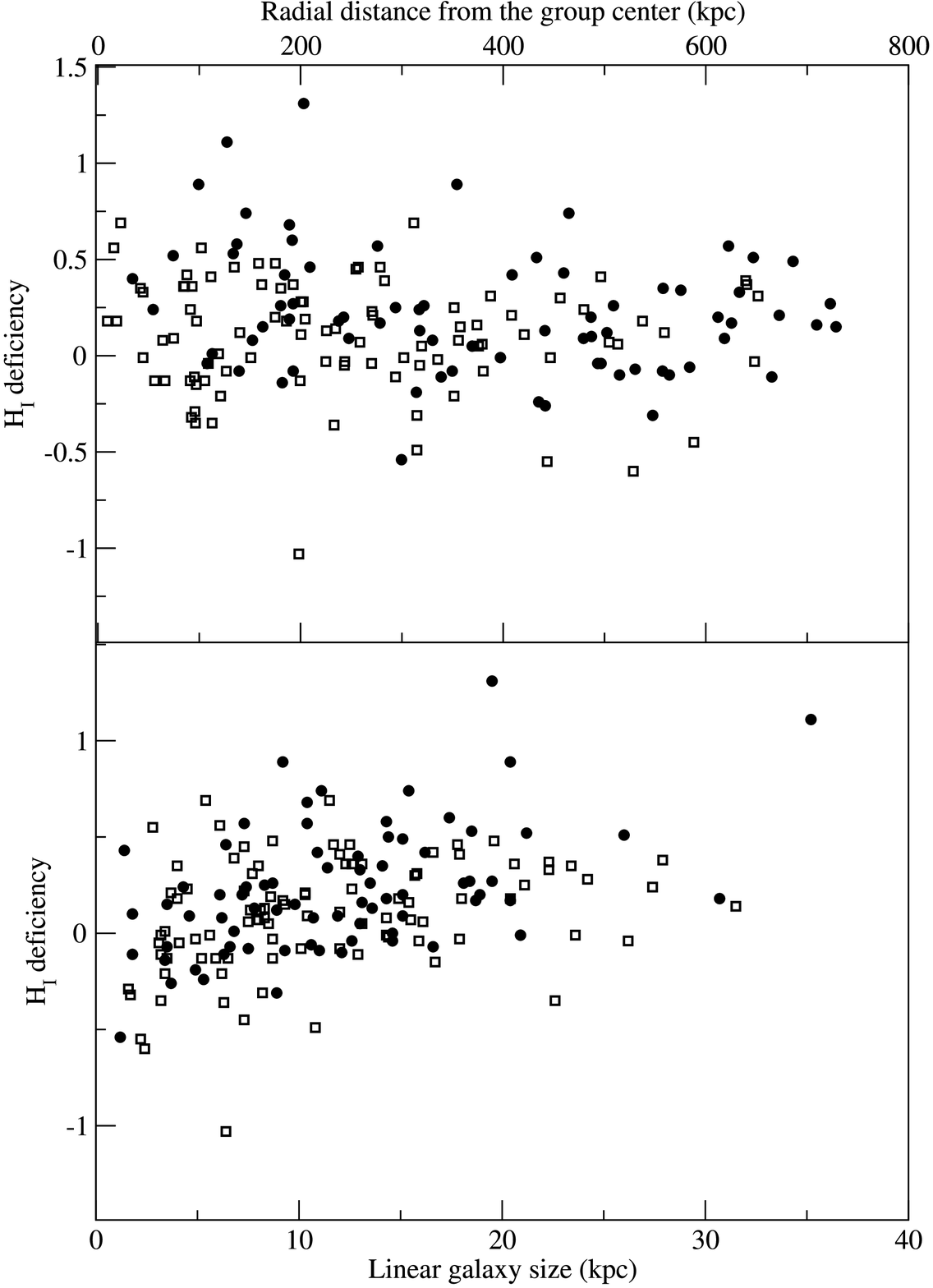}{The top panel shows the distribution of \HI ~deficiency with respect to the distance from the group centre. The bottom panel shows the distribution of \HI ~deficiency with respect to the linear sizes of the galaxies. The number of galaxies in groups with and without Xray emission are 74 and 96, respectively.}{5.1}{1fig}{-0.2}{0}{-0.0}
Both local and large scale environments affect the \HI ~content in galaxies. Interactions and mergers are processes that are likely to be similar in groups with and without diffuse X-ray emission.
Therefore, we first tested if the presence of diffuse X-ray gas affects the \HI ~content any further. We find that, on average, the galaxies in X-ray groups have a \HI ~deficiency of 0.28 $\pm$0.04 whereas the ones in the non X-ray groups show an insignificant deficiency of 0.09 $\pm$0.03. A few groups in the later catagory, eg. NGC584, NGC7582 and UGC9858, that have significant \HI ~deficiency also have upper limits to their X-ray luminosities close to the lowest luminosity among the X-ray detected groups.

For the purpose of this analysis, we constituted two {\it composite} groups, one with and another without diffuse X-ray emission, using all the \HI~ detected group members. The group distances are derived from the mean radial velocities of the \HI ~detected members. These are used to convert the projected angular distances of the group members with respect to group centres to linear distances and the optical major diameters, d$_{l}$, to linear sizes. The top panel of Fig.~1 shows \HI ~deficiency, as defined in section 2, 
for all members of these {\it composite} groups with respect to their radial distances. No variation of \HI~ deficiency 
with radial distance from the group centre is seen, but the fractional number of \HI~ deficient galaxies in X-ray groups is higher than in non X-ray groups. The bottom 
panel of Fig.~1 shows \HI ~deficiency 
distributed with respect to the linear sizes of the galaxies. 
\HI ~deficiency seems to increase with size which could be a bias. \mh1 is observed to be proportional to D$_{l}^{1.7}$ \citep{H1Def}. Thus \HI ~surface density decreases with D$_{l}$ as  D$_{l}^{0.3}$. So computing \HI ~deficiency with a constant value of \smh ~for each type leads to an overestimate for galaxies of large diameters and an underestimate for those with small ones. The effect present in Fig.~1 agrees quantitatively with this explanation. The average effect of such a bias on the galaxies of the sample leads to an underestimate of 0.04 on \hdef. However, this does not affect the results of our comparative study.

\cfig{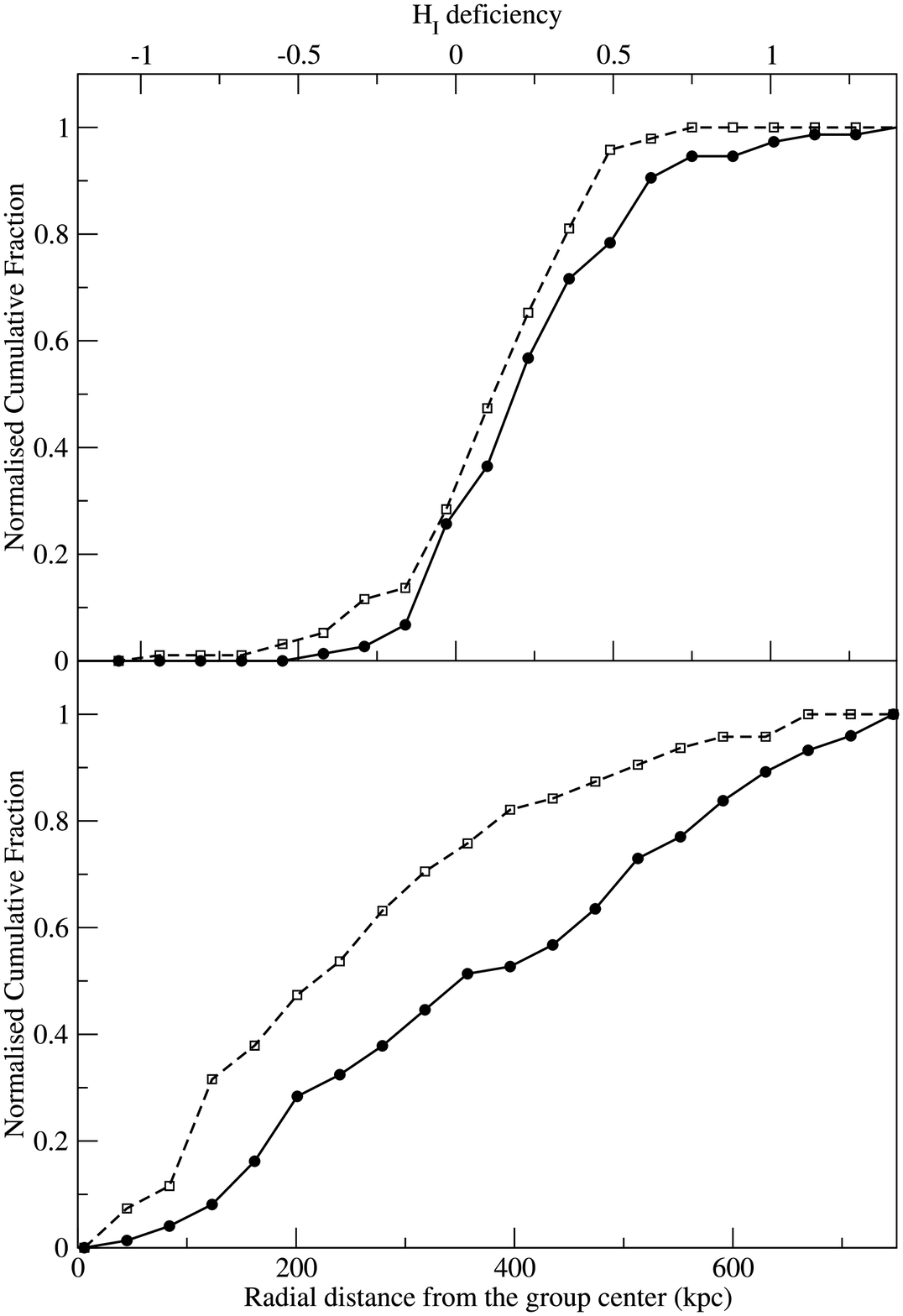}{The top panel shows the behaviour of normalised cumulative fraction of the two sets of galaxies with respect to the \HI ~deficiency; the bottom panel shows the same with respect to the radial distances from their group centres. Filled circles connected by solid lines represent galaxies in X-ray bright groups whereas empty squares connected by dashed lines represent galaxies in non X-ray groups}{4.9}{2fig}{-0.0}{0}{-0.2}
The top panel of Fig.~2 shows the cumulative distribution of \HI ~deficiency for the two categories, Xray-bright and non X-ray groups. Cumulative distribution represents, for each value, q$_{i}$, of the quantity, q, given in the abscissae, the proportion of galaxies, for which q $<$ q$_{i}$. Three statistical tests, Kolmogorov-Smirnov (KS), Mann-Whitney (MW) and Wald-Wolfowitz (WW), were performed on the cumulative distributions of the quantity to test if the two samples are drawn from the same parent population. In the case of \HI ~deficiency the probabilities are: KS: 8.0\%; MW: 6.0\% and WW: 1.1\%. The low probabilities indicate that the two groups are unlikely to be drawn from the same parent distribution, i.e. one group tends to be more \HI~ rich relative to the other.  
Small number statisitics, projection effects and distance uncertainties 
may still affect these results to some degree. Nonetheless, the above results support the premise that the galaxies in X-ray bright groups have statistically lower \HI~ content and thereby higher \HI~ deficiency compared to the ones in non X-ray groups. 

The bottom panel of Fig.~2 shows the cumulative distribution of radial distances from the group center for the two categories, Xray-bright and non X-ray groups. The same three tests were performed on these distributions. Probabilities that the two composite groups are drawn from the same parent distribution from KS, MW and WW tests are 0.08$, \%$ 0.01$\%$ and 33.6$\%$, respectively. Clearly, the X-ray bright groups are more extended compared to the non X-ray bright groups. The average radius of the X-ray bright groups is 600 kpc while that of the non X-ray bright groups is 475 kpc.

\cfig{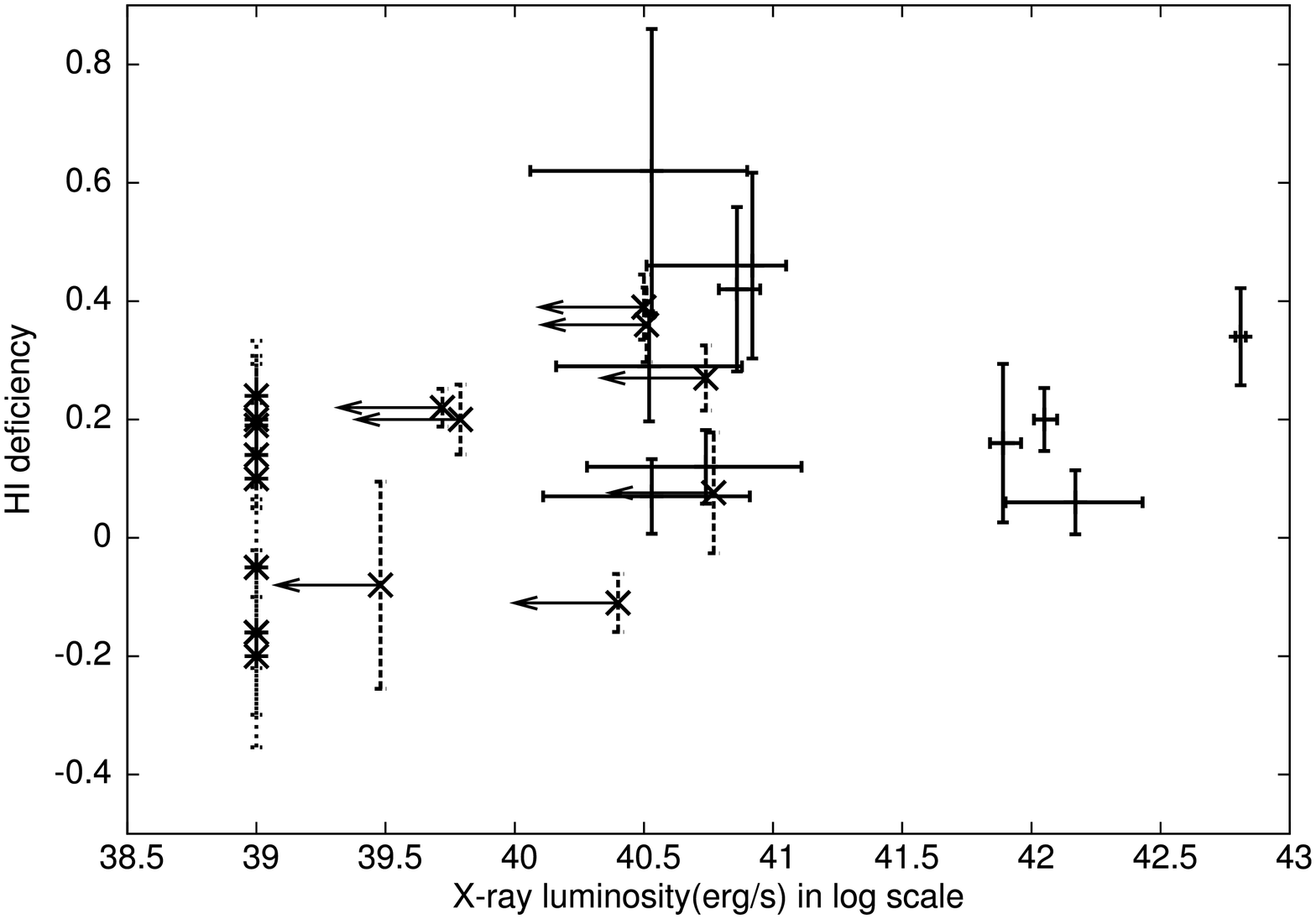}{Variation in \HI ~deficiency with L$_{x}$ is shown.}{2.3}{defLx}{-0.1}{0}{-0.0}

Fig.~3 shows the plot of the group deficiency, as defined in section 2, against the X-ray luminosity (in log scale) of the studied groups. In this figure, the \HI~ deficits of the nine all-spiral groups  for which the X-ray luminosities are unknown, are marked at an X-ray luminosity ($erg ~s^{-1}$ in log scale) of 39.0. X-ray bright groups are represented with $\pm ~1 ~\sigma$ errorbars on both X and Y values; X-ray non-detected groups having upper limits in X-ray luminosity are shown with arrows pointing to the left.
No dependence with X-ray luminosity is seen, excepting that X-ray bright groups are deficient and non X-ray groups have near normal \HI ~content.
 
\cfig{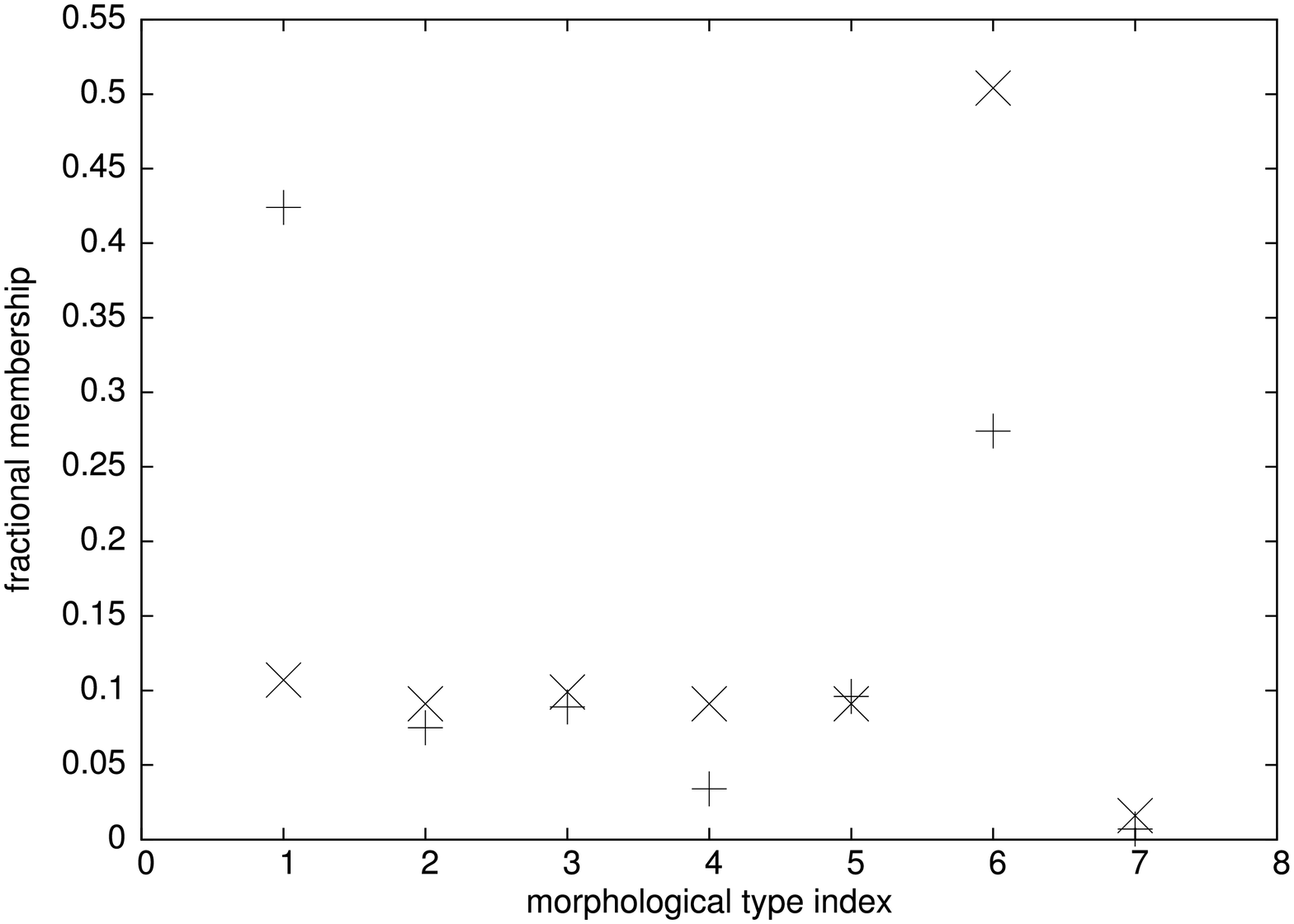}{Distribution of galaxies from groups with and without diffuse X-ray emission with respect to morphological type index from Table 1 (here, Elliptical, S0 and S0/a types are grouped under index 1). Symbol `+' represent X-ray bright groups and `x' represent non X-ray groups.}{2.3}{MpT}{-0.1}{0}{-0.0}

The fraction of galaxies according to the morphological type index for the two categories are shown in Fig.4. As noted in other instances (Mulchaey \etal, 2003), we also find that the X-ray bright groups have a larger fraction of early type galaxies compared to non X-ray groups. Interestingly, the early type members are preferentially located close to the centers in the X-ray bright groups (Tables 3 \& 4). Their exclusion in our analysis and the larger membership of the X-ray bright groups can possibly explain the finding that the X-ray bright groups are more extended than the non X-ray groups.

\section {Discussion}
Environment affects substantially the gas content, star formation rate and morphology of a galaxy as
demonstrated by studies of cluster galaxies. Some observed properties of clusters viz. the population evolution with radius
ie. from red, evolved and early type in the inner parts to bluer, younger and later type in the outer parts,
the increase in the fraction of blue-galaxies ({\it Butcher-Oemler effect}; \citealt{BuOem}) and decrease in
the fraction of S0s ({\it morphological evolution}) with redshift, have led to
considerations of mechanisms that change the star formation rate and morphology of cluster galaxies.
Removal of the warm gas from the halo ({\it strangulation}) and the cold-gas from the disk ({\it
ram pressure stripping} and {\it evaporation via thermal conduction}) have been invoked to change the star
formation rate. Major and minor mergers have been invoked to account for the {\it morphological evolution} (eg. \citealt{MorEv1}, \citealt{MorEv2}).
However, similarity in spectral and morphological properties of poor and more massive clusters at a redshift
of $\sim$0.25 \citep{PreProc1} 
and the level of suppression of star formation in galaxies in many clusters
even at a radius of $\sim$1 Mpc are taken to imply {\it pre-processing} in {\it sub-clusters}: ie. some of
the warm gas from the halo and the cold gas from disk seem to be lost in processes in less massive clusters
at a higher redshift that merged to form today's rich clusters. {\it Strangulation, ram pressure} and
{\it evaporation} seem to be playing a role in them \citep{PreProc2}. 

 Interactions and mergers are processes that are likely to be similar in groups with and without diffuse X-ray emission. Besides, tidal interactions are rare in group environments: for typical membership number (6 spirals), size (1.5 Mpc diameter) and dispersion (150 km/s) of a group, the number of encounters for 50\% gas loss is 0.03 for the group (0.005 per spiral galaxy) in Hubble time \citep{chamaraux}. Therefore, the amount of gas lost in this way seems insufficient to explain the observed group deficiencies in X-ray bright groups. However, our study suggests that processes such as {\it tidal aided ram pressure stripping} 
\citep{TRamP} 
and {\it evaporation via thermal conduction} may be effective even among present day loose groups. Our findings are: (a) the groups with diffuse X-ray emission seem to have lost  more gas
compared to groups without diffuse X-ray emission; (b) 
the X-ray bright groups are spatially more extended than the non X-ray groups. These findings indicate a role for the X-ray emitting gas in aiding \HI~ removal from the galaxies.

In cluster environment ram pressure stripping has been seen to be an effective process for removing gas from galaxies. In groups this process was not considered to be an efficient one as lower IGM density and velocity dispersion, both smaller by an order of magnitude compared to clusters, make direct ram pressure
stripping effective only below a critical \HI~ column density of $10^{19}$ per $cm^{2}$ for a normal galaxy with an optical radius of 10 kpc, and $10^{11}$ stars . However in X-ray bright groups this picture can be a little different. From the quoted X-ray luminosity and temperature in \cite{DifX},
we have calculated the IGM densities assuming the emission to originate from thermal bremsstrahlung. The IGM densities thus calculated for 10 X-ray bright groups vary from 5$ ~\times~ 10^{-4}$ to 2$ ~\times~ 10^{-3}$. For a normal galaxy with an optical radius of 10 kpc and $10^{11}$  to $10^{10}$ stars, and for typical velocity dispersion quoted in the literature, the ram pressure effects become important at \HI~ column densities $\sim$$10^{19}$ to $\sim$$10^{20}$ per $cm^{2}$. The range in stellar mass reflects the range in surface matter density. Ram pressure becomes effective even at higher \HI~ column densities for low surface brightness galaxies.
Low surface brightness galaxies do form a substantial fraction in groups.
The velocity dispersion of galaxies in these groups are also not well determined. Recent discovery of several new members in NGC5044, an X-ray bright group, indicates that some of these groups can actually be more massive. In NGC5044, the new membership has increased its velocity dispersion from 119 km/s to 431 km/s \citep{cellone}. 
For such a velocity dispersion and for a normal galaxy (with an optical radius of 10 kpc, and $10^{11}$ to $10^{10}$ stars), the ram pressure effects are significant at \HI~ column densities of $\sim~ 5~ \times~ 10^{19}$ to $\sim~ 5 ~\times~ 10^{20}$ per $cm^{2}$.

 We have calculated the maximum gas loss possible through ram pressure alone for two galaxies in NGC 5044 group. We have assumed a velocity dispersion of 431 km/s and an IGM density of 4 particles/cc. The later is about half the IGM density determined using X-ray data for this group \citep{DifX}.
We chose two galaxies differing in stellar mass by a factor of ten to determine the varying effects of ram pressure on different types of galaxies in a group. We determined their stellar masses using their J and K magnitudes \citep{Bell}.
In the first case of MCG-03-34-04, the stellar mass is found to be $\sim 1.5~\times ~10^{10}$ M$_{\odot}$, and the corresponding critical \HI ~column density beyond which the ram pressure can strip off gas is $\sim~4~\times ~10^{19}$ per $cm^{2}$. For a gaussian distribution of \HI ~\citep{chamaraux} and using a peak column density of $3.2 ~\times ~ 10^{21}$ per $cm^{2}$ (from our interferometric data), it is seen that $\la$ 5\% of the gas can be stripped off in this way. In the second case, MCG-03-34-041 is found to have a stellar content of $\sim~4~\times ~10^{9}$ M$_{\odot}$ and a peak column density of $2.6 ~\times ~ 10^{21}$ per $cm^{2}$, leading to a gas loss of 32\%.  


Thus, ram pressure is probably an important gas removing mechanism in X-ray bright groups, either on its own or assisted by tidal interaction that stretches the gas below the critical column densities. However this is a density dependent mechanism and thus will be less efficient as the galaxy recedes the group centre, since the IGM density will decrease with increasing distance from the centre. Also for groups with smaller velocity dispersion, this process will become much less effective as the ram pressure depends on the square of the velocity dispersion. For example, a galaxy like MCG-03-34-041 will lose only 5\% gas if the velocity dispersion is 150 km/s. It may also be noted that for the ten X-ray bright groups for which we could determine the IGM densities ($\rho$), the quantity, $\rho ~\sigma_{\rm v}^{2}$, is found to have no correlation with the group deficiency.

Evaporation via thermal conduction seems to be another process which can be responsible for mass loss from galaxies embedded in a hot medium \citep{Evap}. 
It is interesting to note that {\it evaporation} depends directly on temperature and weakly on density, complementary to ram pressure processes. At temperature of 5$ ~\times~ 10^{6}$ K, typical of an X-ray bright group, a disk galaxy can loose as much as 4$ ~\times~ 10^{7}$ M$_{\odot}$ in the time it takes to cross the central 100 kpc region at 200 km/s speed. To expell a mass of 10$^9 ~M_{\odot}$ over a period of 10$^{9}$ years, it requires a mechanical power input of about 10$^{39.5} ~erg ~s^{-1}$, a fraction of the luminosity of the X-ray bright groups. This suggests that if a small fraction of the thermal energy in the hot gas is coupled to the galactic cold gas over a billion years, the gas in the outer parts of the galaxies can be stripped leading to the observed deficit in \HI. The saturation of conductivity and effects of magnetic field may reduce this mass loss. However, mixing from gas-dynamic instabilities as the disk ploughs through the tenuous IGM may enhance the mass loss. Evaporation may also become {\it asymmetric} caused by the galaxy motion and the gradient in the temperature and density of the hot gas. Such asymmetric {\it evaporation} may further deposit momentum to the adjoining gas via (Spitzer's) rocket effect and enhance the gas loss. Numerical simulations of this process taking into account such effects may yield quantitative results and verify the viability of this mechanism. Such an attempt is beyond the scope of this paper. 

We plan to enlarge the data set by obtaining single dish data on more groups, with and
without diffuse X-ray emission. This will help us to improve the confidence level in these results. We are also
in the process of obtaining high resolution images of some members from both kinds of groups, looking for
morphological evidence to try to pin down the responsible process. Extended, asymmetric, low surface brightness \HI ~distributions are typical signs of tidal interaction. Swept-back appearence of the \HI ~gas along with the asymmetric structures, would suggest {\it tidal aided ram pressure stripping}. Evaporation will result in galaxies with somewhat smaller than usual \HI~diameters, and a depressed \HI~surface density across the entire face of the galaxy \citep{Evapc}. 
Thus, with sensitive synthesis data at moderate resolution, we hope to be able to assess the nature of gas removal and possibly conclude on the viable mechanism. 

\section {Conclusions}
We have studied the \HI ~content of galaxies in loose groups with and without diffuse X-ray emission. We find that the galaxies in non X-ray groups are not deficient in \HI ~with respect to the field galaxies. The galaxies in X-ray groups are clearly deficient in \HI ~and have lost more gas (\HI) compared to those in non X-ray groups. 
No systematic dependence of the \HI ~deficiency with $L_{x}$ is found.
We also find that the X-ray groups are more extended than non X-ray groups. 
{\it Tidal aided ram pressure stripping} and {\it evaporation} are the possible mechanisms leading to the excess gas loss found in galaxies in X-ray groups.

{\begin{table}
 \caption{Details of the galaxies in groups with diffuse X-ray emission}
 \begin{tabular}{p{2.2cm} l l r r r r}
\hline
Galaxy &{\it MT} &$d_{l}$(\amin)  &{\it SMD} & \hdef &Ang. & Tel. \\
\hline
\multicolumn{7}{c}{\bf NGC524 (25.5 Mpc)}\\  
\hline
NGC0524 &S0 &2.8 &-  &- &0.0 &-\\
NGC0516 &S0 &1.4 &-  &- &9.8 & - \\
NGC0518 &Sa &1.7 &6.81  & $-$0.04 &14.5 &H \\
NGC0532 &Sab &2.5  &6.24  &0.53 &18.0 &H \\
NGC0509 &S0  &1.6 & - & -&21.6 &- \\
IC0101 &S? &1.4 & 6.23 & 0.68 &25.4 &A\\
NGC0522 &Sc &2.63 &5.56  &1.31 &27.4 &A \\
IC0114 &S0 &1.7 & - &-  &32.3 &-\\
CGCG411-038 &S0 &0.6 &-  &- &36.9 &-\\ 
NGC0502 &S0 &1.1 &-  &- &40.4 & -\\
NGC0489 & S0  & 1.7 & - &- &47.3 & - \\ \hline
\multicolumn{7}{c}{\bf NGC 720 (16.6 Mpc)} \\ \hline
NGC720 & E   &4.7 &-&- &0.0 &- \\
2MASX-1$^{a}$ & S0 & 0.7  &-&-&14.8 &-  \\ 
2MASX-2$^{b}$ & Sc  &0.9 & -&- &32.3 &X \\ 
MCG-02-05-072  & S0/a & 1.3  &-&- &34.4 &-  \\
KUG0147-138 & Sb &1.5 &6.70&0.20 & 50.2& H \\
DDO015 & Sm &1.9   &6.06&0.89 &73.4 &H \\
ARP004 & Im  &2.8  &6.69&0.26 &105.4 & H\\
UGCA022 & Sdm  &2.7  &6.62&0.33 &131.1 & H\\
\hline
\multicolumn{7}{c}{\bf NGC 1589 (37.8 Mpc)} \\
\hline
NGC1587 &Epec  & 1.7 &- &-&0.4 &- \\
NGC1588 &Epec  & 1.4 &- &-&0.8 & -\\
NGC1589$^{+}$(c) & Sab & 3.2&5.66&1.11 &11.9 &H \\
UGC03072$^{+}$ & Im  & 1.3 &6.37&0.58 &12.8 &A \\
UGC03058 & Sm  &1.3&6.77&0.18 &17.6 &H  \\
NGC1593 & S0  &  1.6   &6.44&- &22.2 & -\\
UGC03080 & Sc  & 1.9  &6.88&$-$0.01 &37.0 &H \\
UGC03054 & Sd  &1.4&6.21&0.74  &43.3 &H \\
NGC1586 & Sbc  &1.7 &6.76&0.17 &58.3 &H \\

\hline
\multicolumn{7}{c}{\bf NGC3686 (12.2 Mpc)}\\  
\hline
NGC3607  & S0  &4.9  &- & -&0.0 &-  \\
NGC3608 &E &3.2&-&-&5.7 &- \\
UGC06296 & Sc  & 1.2  &6.63 & 0.24&15.4 &A \\
LSBCD570-04 &dI & 1.0  &- &- &29.9 & -\\
UGC06341 & Sdm& 1.0  &6.80 &0.15 &46.0 &A \\
UGC06324 &S0 & 1.7  &- &- &46.0 &- \\
NGC3626 &S0 &2.7   &- & -&48.5 &- \\
UGC06320 &S?  &0.95   &7.05 &$-$0.14 &51.4 &A \\
NGC3592 &Sc & 1.8  &6.41 & 0.46 &59.2 & A\\ 
LSBCD570-03&dI &0.8   &- &- &93.7 & -\\
NGC3659  & SBm &2.1 &7.03 &$-$0.08 &98.9 &A \\
UGC6300&E&1.2&-&-&103.8 & -\\
NGC3655 &Sc  &1.5   &7.11 &$-$0.24 &123.0 &G \\
LSBCD570-02  &Im &0.39   &6.52 &0.43 &130.0 &A \\
LSBCD570-01 &Sm & 0.5  &6.85 & 0.10 &137.7 &A \\
UGC06171 &IBm &2.5   &6.83 &0.12 &142.0 & RC3 \\
NGC3681 & Sbc &2.5   &7.24 &$-$0.31 &154.8 & G \\
UGC06181 &Im &1.0   &7.02 &$-$0.07 &157.5 & G \\
NGC3684 &Sbc  & 3.1  & 7.02 &$-$0.09 &159.4 &N91  \\
NGC3686 &Sbc &3.2   &6.59 & 0.34&162.7 &N91 \\
NGC3691 & SBb& 1.3 &6.82  &0.09 &174.8 & A \\

\hline
\end{tabular}
\end{table}
}

{\begin{table}
 \contcaption{}
 \begin{tabular}{p{2.2cm} l l r r r r}
\hline
Galaxy &{\it MT} &$d_{l}$(\amin)  &{\it SMD} & \hdef &Ang. & Tel. \\
\hline
\multicolumn{7}{c}{\bf NGC 4261 (21 Mpc)} \\
\hline
NGC4261 &E &4.1 & - &-&0.1 &-  \\
VCC0344 &E &.34 & - &- &1.9 & - \\
IC3155 &S0 &1.17 &-  &- &12.0 & - \\
VCC0292 & dE &0.35 & - &- &13.0 & - \\
NGC4269 &S0+ &1.1 & - &- &13.0 &-  \\
VCC0388& dE &0.51 &- &- &14.9 &-  \\
NGC4260&Sa &3.34 &5.87  &0.89  &16.3 & B \\
VCC0405&dE &.18 & - &- &17.1 & - \\
NGC4287&S &1.82 & 6.17 &0.74 &23.9 & H\\
VCC0332&S0 &0.29 & - &- &28.9 & - \\
NGC4277&S0/a &1.13  &-  & -&30.8 & - \\
VCC287 &dE&0.51&-&-&37.3&-\\
VCC0223 &BCD? &0.2 & 7.68  &$-$0.54&49.1 & A  \\
VCC0297 &Sc  &1.01 & 6.79 & 0.08 &54.1 &H  \\
VCC238&dE&0.25&-&-&59.3&-\\
NGC4223 &S0 &2.6 &-  &- &59.5 & - \\
HARO06 &E &0.49 & - &- &61.1 & - \\
NGC4215 &S0 &1.9 & - &-&62.3 & -  \\
NGC4255 &S0 &1.3 & - &- &62.8 &-  \\
UGC07411 &S0/a &1.4 & - & -&70.7 &-  \\
NGC4197$^{+}$(c) &Sc  &4.26  &6.36  &0.51 &70.9 &H  \\
VCC0114$^{+}$ &Im &0.6 &7.21  &$-$0.26 &72.3 & A \\
NGC4292 &S0 &1.7 &-  &- &79.0 & - \\
VCC0693 &S? &1.0 &6.71  &0.20 &79.7 & A \\
VCC0256 &S  &0.70 & -  & -&84.4 & - \\
VCC0172 &Im &1.08 &7.02  &$-$0.07 &86.8 & B \\
VCC764&S0 &0.54 &-&-&90.2&-\\
VCC468&BCD&0.3&7.25&$-$0.11&108.9&A\\
NGC4233&S0&2.4&&-&-113.0&-\\
NGC4180&Sab&1.6&6.61&0.15 &119.3&A\\
\hline
\multicolumn{7}{c}{\bf NGC 4589 (16.7 Mpc)} \\ 
\hline
NGC4319&Sab &3.0 & - & -&11.2 &ND\\
NGC4386 &S0 &2.5 & - & -&11.8 &-\\
NGC4291 &E &1.9 & - &- &16.8 &-\\
NGC4363 &Sb &1.4 &6.90  &0.01 &23.3 &B\\
NGC4331 &Im &2.2 &6.86  &0.08 &51.0 &G\\
UGC07189 &Sdm &1.7 &6.70  &0.25 &60.5 &B\\
UGC07265 &Sdm &1.0 & 7.14  &$-$0.19 &64.7 &G\\
UGC07238 &Scd &1.53  &6.71  &0.24&65.3 &B \\
UGC07872 &Im  &1.6 &6.81   &0.13 &65.4 &B\\
NGC4133 &Sb  &1.8 &6.65  &  0.26&66.2 &N91\\
NGC4159 &Sdm   &1.3 &7.06  &$-$0.11&69.8 & B\\
NGC4589 &E &3.2 & - & -&84.8 &-\\
NGC4648 &E &2.1 & - &- &86.2 &-\\
NGC4127  &Sc &2.5 & 6.97 &$-$0.10 &106.0 &RC3\\
UGC7844 &Sd&1.27&-&-&116.3&X\\
UGC7908 &Scd&1.5&6.38&0.57&128.1&G\\
UGC7086  &Sb&2.69&6.74&0.16&146.0&G\\
\hline
\end{tabular}
\end{table}
}

{\begin{table}
 \contcaption{}
 \begin{tabular}{p{2.2cm} l l r r r r}
\hline
Galaxy &{\it MT} &$d_{l}$(\amin)  &{\it SMD} & \hdef &Ang. & Tel. \\
\hline
\multicolumn{7}{c}{\bf NGC 5044 (26.0 Mpc)}\\ 
\hline  
NGC5044& E &3.0 &-&-&0.5 &\\
NGC5044-1$^{c}$&dE&0.4&-&-&5.3 &\\                        
NGC5049 & S0  & 1.9  &-& -&8.0 &\\
LEDA083813&dE&0.4&-&-&19.0 &\\ 
NGC5030 & S0 &  1.8 & - &-&23.0 &\\
MCG-03-34-041& Sc  &2.3  &6.27&0.60&25.3 &H \\
NGC5031 & S0  &1.6 &- &- &25.3 &-\\ 
LEDA083798&Sd&0.6&-&-&28.5 &-\\
LCSBS1851O&dS0&0.7&-&-&32.6 &-\\
MCG-03-34-020  & E &0.6&-&- &35.5 &-\\
NGC5017 & E & 1.8 &-&-&43.0 &-\\
IC0863 & Sa  &1.8 &6.63&0.13 &58.1 &H\\
UGCA338 & Sdm  & 2.0 &6.86&0.09&63.1 &H \\
MCG-03-34-014 & Sc  &2.5 &6.67&0.20 &80.6 &H\\
MCG-03-34-004 & S0  &1.9 &- & -&83.0 &- \\ 
SGC1317.2-1702 & Sdm &1.9  &6.44&0.50 &85.2 &H \\
SGC1316.2-1722 & Sm  & 2.0 &6.46&0.49 &90.4&H \\
\hline
\multicolumn{7}{c}{\bf UGC12064 (50.2 Mpc)} \\
\hline
UGC12064(c)  &S0 & 1.1 & - &- &0.2 &-\\
UGC12073(c) &Sb &2.1  &6.73& 0.18&16.3 &G\\
UGC12075(c) &Scd &1.4 & 6.77 &0.18 &19.1 &G\\
UGC12077(c) & S  &1.0 &6.95  & $-$0.04&33.8 &G \\
UGC12079(c) &Dwarf &1.0 &6.99  &$-$0.04 &34.0 &G\\

\hline
\multicolumn{7}{c}{\bf IC 1459 (17.8 Mpc)}\\  
\hline
IC1459 &E &5.2&- &-&0.1 &-\\
IC5264$^{+}$(c) & Sab  & 2.5  &6.37 &0.40&6.6 &P\\
IC5269B & Scd  &4.1  & 6.43 &0.52&14.3 &H\\
IC5269(c) & S..  &1.8   &7.0 &$-$0.08& 26.9&H\\ 
NGC7418 & Scd  &3.5  &6.69 &0.26& 34.9& H\\
IC5270(c) & Sc  &3.2   &6.94 &$-$0.08&37.2 &H\\
NGC7421 & Sbc  &2.0   &6.36 &0.57&53.3 &H\\
ESO406-G031  & Sb &1.5   &- &-&64.1 &ND\\
IC5269C & Sd  &2.1  &6.53 &0.42&79.0 &H\\
\hline
\multicolumn{7}{c}{\bf NGC7619 (37.2 Mpc)} \\
\hline 
NGC7619 & E  &2.5 &-  &- &0.1 &-\\
NGC7626 &E &1.3 & - &- &6.8 &-\\
UGC12510&E&1.3&- & -&9.4 &-\\
NGC7623 &S0 &1.2 &-  &- &11.9 &-\\
NGC7611 &S0 &1.5 & - &-&12.8 & -\\
KUG2318+078 &S? &1.1 & 6.82 &0.09 &14.1 &A\\ 
KUG2318+079B &Sc &0.3 &-  &- &16.2 &X\\
NGC7608 &S? &1.5 &6.49  &0.42&17.0 &A\\
NGC7631 &Sb &1.8 &6.64  &0.27&17.8 &A\\
NGC7612 &S0 &1.6 & - &- &23.4 &-\\ 
UGC12497 &Im &1.2 &6.90  &0.05&34.1 &A\\
NGC7634  &S0 &1.2 &-  & -&46.1 &-\\
CGCG406-086 &S  &1.3 &6.56  &0.35&51.5 &A\\
FGC284A  &Sc &1.12 & - & -&51.7 &X\\
UGC12480 &Im &.98 &7.00  &$-$0.06&54.0 &A\\
NGC7604 &E  &0.3& - &- &58.6 &-\\
UGC12561 &Sdm &1.4 &6.74  &0.20&62.1 &A\\
UGC12585 &Sdm &1.7 & 6.67 &0.27&66.8 &A\\

\hline
\end{tabular}
Blue compact dwarfs in NGC4261 group are denoted by BCD for their morphological type and are taken to belong to the category of peculiars. \\ 
$^{a}$	2MASXJ01535632-1350125, \hfill $^{b}$  2MASXJ01524752-1416211, \\ 
$^{c}$  NGC5044 GROUP:[FS90] 076; \hfill $^{+}$ Please see Table 4 footnote.
\end{table}
}







{\begin{table}
 \caption{Details of the galaxies in groups without diffuse X-ray emission}
 \begin{tabular}{p{2.2cm} l l r r r r}
\hline
Galaxy &{\it MT} &$d_{l}$(\amin)  &{\it SMD} & \hdef &Ang. & Tel. \\
\hline
\multicolumn{7}{c}{\bf NGC 584 (18.7 Mpc)}\\
\hline
NGC596&E+pec &3.2 &-&- &7.3 &- \\
NGC600&Sd  &3.3 &6.77&0.18 &17.9 &H \\
NGC586&Sa  &1.6 &- &- &27.1 &ND \\
KDG007(c)&Dwarf &1.6 &6.47&0.48 &29.1 &H \\
NGC584&E  &4.2 & -& -&31.3 & -\\
NGC615(c)&Sb &3.6 & 6.43&0.48 &32.2 & H \\
IC0127&Sb   &1.8 &- &-&53.1 &ND \\
NGC636&E  &2.8 &- &- &90.5 &- \\
UGCA017&Sc   &2.9 & 6.56&0.31&119.8 & H \\
\hline
\multicolumn{7}{c}{\bf NGC628 (7.4 Mpc)}\\ 
\hline
NGC628&Sc&10.5&7.22 & -0.35&44.7 &GRA\\
UGC1171&Im&1.3&6.39&0.55&47.4 & A\\
UGC1176&Im&4.6&6.61 &0.33&50.3 &G \\
KDG010&Dwarf&1.5&-&-&80.8&X\\
IC0148&Im&2.95&7.98&$-$1.03&92.1 &G \\
NGC0660&Sa&8.3&6.80&$-$0.03&113.2 &N91 \\
UGC01200&Im&1.5&7.06&$-$0.11&136.5 &A \\
UGC01104&Im&1.0&7.50&$-$0.55&206.0 & G\\
UGC01246&Im&1.5&6.96&$-$0.01&207.5 &A \\
UGC01175&Sm&1.1&7.55&$-$0.60&245.5 & G\\
\hline
\multicolumn{7}{c}{\bf NGC 841 (45.1 Mpc)}\\ 
\hline
NGC845 &Sb &1.7 &6.54 &0.37 &12.3 &N \\
UGC1721&Sbc&2.0&6.97&$-$0.04&20.6&A\\
NGC841&Sab&1.8&6.78&$-$0.01&23.0&A\\
NGC834&S?&1.1&6.93&$-$0.02&25.5&A\\
UGC1673&S?&1.0&-&-&28.7&X\\
UGC1650&Sd&2.13&6.56&0.38&48.7&A\\

\hline
\multicolumn{7}{c}{\bf IC1954 (9.2 Mpc)}\\ 
\hline
NGC1311& Sm&  3.0& 6.59&0.35&31.5 &H  \\
IC1933& Sd&  2.2& 7.08&$-$0.13&39.3 & H \\
IC1954& Sc&  3.2& 6.68&0.19 &76.0 & H\\
ESO200-G045 & Im  & 2.0& 6.26&0.69 &116.0 &H  \\
NGC1249& Sdm&  4.9& 6.90&0.05 &139.8  & H\\
IC1959& Sm&  2.8& 6.89&0.06 &141.1 & H\\

\hline
\multicolumn{7}{c}{\bf NGC 1519 (18.8 Mpc) }\\ 
\hline
NGC1519$^{+}$(c)&Sb  & 2.1& 6.22 &0.69&4.2 & H \\
UGCA088$^{+}$&Sdm  & 1.9&  6.86& 0.09 &13.7 &G \\
UGCA087&Sm & 2.3& 6.72 &0.23 &49.4 & H \\
SGC0401.3-1720&Im  & 1.7&  6.80&0.15 &65.3 &H  \\
MCG-03-11-018 &Sm  & 1.4& 6.64 &0.31&70.8 & H \\ 
MCG-03-11-019 &Sdm &1.6&6.98&$-$0.03&118.3&N\\
\hline
\multicolumn{7}{c}{\bf NGC 1792 (11.8 Mpc)} \\
\hline
NGC1808&Sb   &6.5  &6.58&0.33 &0.0 &H  \\
NGC1792 &Sbc  &5.2  &6.47 &0.46&40.5 & H \\
NGC1827 &Scd   &3.0  &6.75&0.20 &43.6 &H \\
ESO305-G009&Sdm  &3.5  &7.03&$-$0.08 & 48.1& H \\
ESO362-G011&Sbc   &4.5  & 6.77&0.16&109.5 & H \\
ESO362-G016&Dwarf   &1.3 & 6.71&0.23 &140.6 &H \\
ESO362-G019&Sm   &2.2  &6.83&0.12&163.0 & H \\
\hline
\end{tabular}
\end{table}

}

{\begin{table}
 \contcaption{}
  \begin{tabular}{p{2.2cm} l l r r r r}
\hline
Galaxy &{\it MT} &$d_{l}$(\amin)  &{\it SMD} & \hdef &Ang. & Tel. \\
\hline
\multicolumn{7}{c}{\bf NGC 2997 (10.6 Mpc)} \\
\hline
ESO434-G030&S(r)&  1.0& -& -&8.2 & ND \\
IC2507(c)&Im& 1.7&  7.08&$-$0.13 &18.3 &H \\
UGCA180(c)&Sm&  2.1& 7.08&$-$0.13 &21.6 &H \\
NGC2997&Sc&  8.9& 6.63&0.24 &29.7 & H\\
UGCA177&Im&  1.1&  6.94&0.01 &38.8 & H\\
ESO434-G041&Im&  1.8&  6.96&$-$0.01&49.2 & H \\
ESO434-G019&Im&  1.3& 6.60&0.35 &58.8  & H \\
ESO373-G020&Im& 1.6&  6.98 &$-$0.03&73.3 & H \\
UGCA182&Im&  2.7&  6.82&0.13&73.5 & H \\
ESO434-G039&Sa-b& 1.0& 6.82 &$-$0.05 &79.3 &H \\
ESO434-G017&dwarf&1.2& 6.74&0.21&88.4 & H \\
UGCA168&Scd&  5.8& 6.54 &0.41& 161.7&H \\
ESO373-G007&Im& 1.3&  6.77&0.18  &175.2 &H \\
\hline
\multicolumn{7}{c}{\bf NGC3264 (9.8 Mpc)}\\ 
\hline
NGC3264&Sdm&2.9&6.87&0.08&22.5 &G \\
UGCA211&pec&0.56&7.43&$-$0.29&33.6 &G \\
NGC3220&Sb&1.23&7.04&$-$0.13&70.0 &N \\
NGC3206&Scd&2.2&7.31&$-$0.36&81.8 &N91\\
UGC05848&Sm&1.45&6.99&$-$0.05&111.4 &G \\
NGC3353&BCD/Irr&1.2&7.35&$-$0.21&123.3 & N43\\

\hline
\multicolumn{7}{c}{\bf NGC 4487 (10.1 Mpc)}\\  
\hline
NGC4504 &Scd  & 4.4 &7.06&$-$0.11 &29.9 &H   \\
UGCA289 &Sdm  & 4.1 &6.54&0.41 &35.0 &H  \\
NGC4487 & Scd  & 4.2 &6.58&0.36&60.4 & H  \\
NGC4597 &Sm  & 4.1 & 6.84 &0.11&131.9 & H\\

\hline
\multicolumn{7}{c}{\bf NGC 5061 (20.8 Mpc)} \\ 
\hline
NGC5061&E  &3.5 &-& - &12.1 & -\\
NGC5078(c)&Sa  &4.0 &6.49&0.27&33.1 & H \\
ESO508-G039&Sm  &1.3 &6.84&0.11  &33.1 &H \\
IC0879(c)&Sab pec &1.2 &6.49 &0.27&33.5 & H \\
IC874&S0  &1.2 & - &-&44.1 &- \\
NGC5101&S0/a &5.4 &- &- &49.5 & - \\  
ESO508-G051&Sdm  &1.4 &6.90&0.05 &52.8 &H \\
IC4231&Sbc &1.7 &6.72&0.21  &67.5 &H \\
ESO508-G059&S..  &1.2 &- &-&93.0 &ND \\
ESO508-G034&Sm  &1.2 &7.40&$-$0.45 & 97.2 &H \\
\hline
\multicolumn{7}{c}{\bf NGC 5907 (7.4 Mpc)}\\ 
\hline

NGC5907&Sc&12.77&6.65&0.22&6.7 &G \\
UGC09776&Im&0.8&7.27&$-$0.32&42.7 &G \\
NGC5866B&Sdm&1.46&7.30&$-$0.35&52.1&G \\
NGC5879&Sbc&3.74&6.80&0.12&64.7 &N91 \\
NGC5866&S0&4.7&-&-&91.7 &- \\

\hline
\multicolumn{7}{c}{\bf UGC 9858 (25.2 Mpc)}\\ 
\hline
UGC09856&Sc&2.26&6.45&0.42&12.0 &G \\
UGC09858&Sbc&4.3&6.79&0.14&32.0 &N91 \\
NGC5929(c) &Sab&1.0&6.31&0.45&34.7 &G \\
NGC5930(c)&Sb&1.7&6.45&0.46&35.0 &G \\
UGC09857(c)&Im&1.6&6.49&0.46&38.0 &G \\

\hline
\multicolumn{7}{c}{\bf NGC 6949 (26.6 Mpc)}\\ 
\hline
NGC6949&S&1.4&7.40&$-$0.49&40.7 & G\\
UGC11613&Sdm&1.85&6.87&0.08&46.0 &G \\
UGC11636&Scd&1.3&7.03&$-$0.08&49.2 & G\\

\hline
\end{tabular}
\end{table}

}

{\begin{table}

 \contcaption{}
 \begin{tabular}{p{2.2cm} l l r r r r}
\hline
Galaxy &{\it MT} &$d_{l}$(\amin)  &{\it SMD} & \hdef &Ang. & Tel. \\
\hline
\multicolumn{7}{c}{\bf NGC 7448 (21.3 Mpc)}\\ 
\hline
NGC 7464&E&0.8&-&-&13.3 &- \\
NGC 7465&S0&1.2&-&-&14.7 &- \\
UGC 12313&Im&1.4&7.08&$-$0.13&15.2 &A \\
NGC 7448&Sbc&2.7&7.08&$-$0.15&16.2 &A \\
UGC 12321&Sbc&1.0&7.14&$-$0.21&20.2 &A \\
NGC 7454&E&2.2&-&-&30.8 & -\\
NGC 7468&E&0.9&-&-&52.0 & -\\
UGC 12350&Sm&2.6&6.89&0.06&85.6 &A \\

\hline
\multicolumn{7}{c}{\bf NGC 7582 (16.1 Mpc)} \\ 
\hline
ESO291-G015$^{*}$(c)&Sa  &1.3 &6.21&0.56 &3.3 &H\\
NGC7582$^{*}$(c)&Sab  &5.0  &6.42&0.35 &9.0 &H\\
NGC7590$^{*}$(c)&Sbc   &2.7  &6.57 &0.36 &18.3 &H\\
NGC7599$^{*}$(c)&Sc  &4.4  &6.51&0.36 &19.9 &H\\
NGC7552&Sab  &3.4  &6.81&$-$0.04 &23.3 &H\\
NGC7632&S0  &2.2  &-&- &41.6 &-\\
ESO347-G008&Sm &1.7  &6.88&0.07 &55.2 &H\\
ESO291-G024&Sc pec   &1.46  &6.48&0.39 &60.4 &H\\
NGC7531&Sbc  &4.5  &6.68 &0.25 &75.1 &H\\
NGC7496&Sbc& 3.3&6.86&0.07&107.7 &H\\
IC5325 &Sbc &2.8 &6.56 &0.36 &136.8 &P\\

\hline
\multicolumn{7}{c}{\bf NGC 7716 (27.0 Mpc)} \\
\hline

NGC7714(c)&Sb pec &1.9 &6.73&0.18&0.0 & H\\
NGC7715(c)&Im pec  &2.6 &6.77&0.18 &2.0 &H\\
UGC12690&Sm  &2.0 &6.65&0.30&58.1 &H \\

\hline
\multicolumn{7}{c}{\bf UGC 12843 (17.6 Mpc)}\\ 
\hline
UGC 12843&Sdm&2.8&6.96&$-$0.01&8.8 &A\\
UGC 12846 &Sm&1.8&6.77&0.17&36.4 &A\\
UGC 12856&Im&1.6&7.26&$-$0.31&61.5 &G\\
MCG+03-01-003&Sm&0.45& -&- &62.0 &X\\

\hline
\end{tabular}
{\bf Symbols used for indicating the telescopes:}

A = Arecibo 1000 ft \hfill
B = Effelsberg 100 m \\
G = Green Bank Telescope 100 m \hfill
GRA = Agassiz Harvard 60ft \\
H = \HI ~Parkes All Sky Survey \hfill
N = Nancay 30 $\times$ 300 m \\
N91 = NRAO 91 m \hfill
N43 = NRAO 43 m \\
P = Parkes 64 m \\
\hrule 
\vspace*{0.05cm}
\hrule 
\vspace*{0.1cm}
The cases of confusion are indicated by a `(c)' in column 1. \\
$^{+}$ HIPASS spectra towards NGC1519, NGC1589, NGC4197 and IC5264 are confused, respectively, with galaxies UGCA088, UGC03072, VCC0114 and NGC7481A. Hence, their deficiencies have been calculated following the prescription mentioned in the text. NGC7481A does not belong to the group as per the criteria and is not in the table. The other three confusing galaxies viz. UGCA088, UGC03072 and VCC0114 have GBT (the first one) or Arecibo (the last two) measurements. Therefore, their deficiencies have been calculated with the higher resolution fluxes and are not marked with a `(c)' in the above table. \\
$^{*}$ NGC7582 is confused with NGC7590 and NGC7599 in one HIPASS pointing and with ESO-291-G015 in another. Since it is closer to the pointing center in the first case, we use the deficiency calculated from that spectrum for this galaxy.
\end{table}

}

\section*{Acknowledgments}
{\noindent The Parkes telescope is part of the Australia Telescope which is funded by the Commonwealth of Australia for operation as a National Facility managed by CSIRO. This research has made use of the NASA/IPAC Extragalactic Database (NED) which is operated by the Jet Propulsion Laboratory, California Institute of Technology, under contract with the National Aeronautics and Space Administration. This research has made use of NASA's Astrophysics Data System. RB thanks his family for their encouragement and support. CS thanks colleagues and friends in the Astronomy group in RRI for useful discussions and suggestions. The authors thank the anonymous referee for useful comments and suggestions.}

\end{document}